\documentstyle[epsf]{l-aa}
\makeatletter
\def\V{Vela X--1}   \def\B{{\em BeppoSAX\/}}
\def\cdof{\mbox{$\chi^2_\nu$}}   
\def\pmt#1#2{_{-#1}^{+#2}}
\def\@cite#1#2{(#1\if@tempswa , #2\fi)}
\def\@citetext#1{\edef\citea@{#1}\expandafter\split@cite \citea@\end@split}
\def\split@cite#119#2\end@split{#1(19#2)}
\def\citetext#1{\if@filesw\immediate\write\@auxout{\string\citation{#1}}\fi
  \@citetext{\@ifundefined{b@#1}{19\@warning
   {Citation(text) `#1' on page \thepage \space undefined}}%
{\csname b@#1\endcsname}}}
\def\newblock{\hskip .11em plus .33em minus .07em}
\makeatother

\title{The \V\ pulse-averaged spectrum as observed by \B}

\author{M. Orlandini\inst{1}
\and D. Dal Fiume\inst{1}
\and F. Frontera\inst{1}
\and G. Cusumano\inst{2}
\and S. Del Sordo\inst{2}
\and S. Giarrusso\inst{2}
\and S. Piraino\inst{2}
\and A. Segreto\inst{2}
\and M. Guainazzi\inst{3}
\and L. Piro\inst{4}
}

\offprints{orlandini@tesre.bo.cnr.it}
\date{Received; Accepted}
\thesaurus{08.02.2; 08.09.2 HD77581; 08.14.1; 08.16.7 \V; 13.25.5}
\institute{Istituto Tecnologie e Studio Radiazioni Extraterrestri (TeSRE),
 C.N.R., via Gobetti, 101, 40129 Bologna, Italy
\and
  Istituto di Fisica Cosmica ed Applicazioni dell'Informatica (IFCAI), C.N.R.,
  via La~Malfa 153, 90146 Palermo, Italy
\and
  \B\ Scientific Data Center (SDC), Nuova Telespazio, Via Corcolle 19, 00131
  Roma, Italy
\and
Istituto di Astrofisica Spaziale (IAS), C.N.R., Via Fermi 21, 00044 Frascati,
  Italy
}

\begin{document}

\maketitle

\begin{abstract}
We report on the 20 ksec observation of \V\ performed by \B\ on 1996 July 14
during its Science Verification Phase. We observed the source in two intensity
states, characterized by a change in luminosity of a factor $\sim 2$, and a
change in absorption of a factor $\sim 10$. The single Narrow Field Instrument
pulse-averaged spectra are well fit by a power law with significantly different
indices. This is in agreement with the observed changes of slope in the
wide-band spectrum: a first change of slope at $\sim 10$ keV, and a second one
at $\sim 35$ keV. To mimic this behaviour we used a double power law modified
by an exponential cutoff --- the so-called NPEX model --- to fit the whole
2--100 keV continuum. This functional is able to adequately describe the data,
expecially the low intensity state. We found an absorption-like feature at
$\sim 57$ keV, very well visible in the ratio performed with the Crab spectrum.
We interpreted this feature as a cyclotron resonance, corresponding to a
neutron star surface magnetic strength of $4.9\times 10^{12}$ Gauss. The \B\
data do not require the presence of a cyclotron resonance at $\sim 27$ keV as
found in earlier works.


\end{abstract}

\section{Introduction}

Among the class of X--ray binary pulsars, \V\ is the one that has been more
carefully monitored since its discovery as a pulsator in 1976 \cite{647}. It is
the prototype of the subclass of pulsars that accrete matter directly from the
stellar wind, coming from the massive early-type star HD77581 (spectral type B0
Ib \cite{734}). During its orbit around the companion, the X--ray emission from
the pulsar suffers substantial photoelectric absorption, increasing until the
source is eclipsed by its companion every 8.96 days --- the orbital period ---
for about 1.7 days \cite{885}. The study of the orbital phase dependence of the
low energy absorption due to the circumstellar matter, together with
spectroscopic studies of the optical companion, has shown the presence of
accretion \cite{88} and photoionization \cite{1345} wakes.

The pulse-averaged spectrum of \V\ also shows an Iron emission line centered at
$\sim 6.4$ keV \cite{303}. Recently, cyclotron resonance features (CRFs) at
$\sim 27$ and 55 keV been reported by \citetext{1538} and \citetext{407}.

\section{Observation}

The \B\ satellite is a program of the Italian Space Agency (ASI), with
participation of the Netherlands Agency for Aerospace Programs (NIVR), devoted
to X--ray astronomical observations in the broad 0.1--300 keV energy band
\cite{1530}.  The payload includes four Narrow Field Instruments (NFIs) and two
Wide Field Cameras \cite{1534}. The NFIs consist of Concentrators Spectrometers
(C/S) with 3 units (MECS) operating in the 1--10 keV energy band \cite{1532}
and 1 unit (LECS) operating in 0.1--10 keV \cite{1531}, a High Pressure Gas
Scintillation Proportional Counter (HPGSPC) operating in the 3--120 keV energy
band \cite{1533} and a Phoswich Detection System (PDS) with four scintillation
detection units operating in the 15--300 keV energy band \cite{1386}.

During the Science Verification Phase a series of well known X--ray sources
have been observed in order to check the capabilities and performances of the
instruments aboard \B. \V\ is one of these sources and it has been observed by
three of the four NFIs (LECS was not operative during this pointing) on 1996
July 14 from 06:01 to 20:54 UT. The net exposure time for MECS, HPGSPC and PDS
was 21.6, 8.1 and 9.5 ksec, respectively. These differences are due to rocking
of collimators (see below) and different filtering criteria during passages in
the South Atlantic Geomagnetic Anomaly (SAGA) and before and after Earth
occultations.

\begin{figure}
%\picplace{10cm}
\epsfxsize=0.5\textwidth   \epsffile{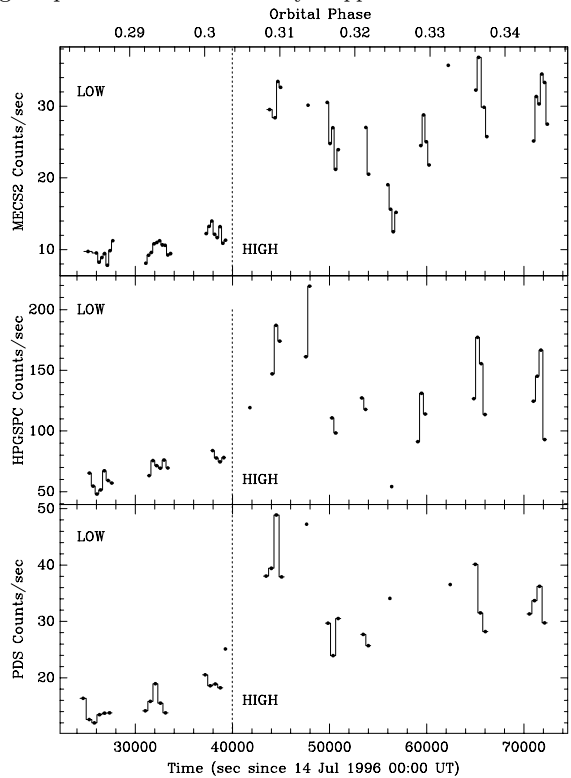}
\caption[]{Vela X-1 background subtracted light curve for all the three
operative \B\ NFIs instruments. Gaps are due to South Atlantic Anomaly passages
and Earth occultations (the average \B\ duty cycle is about 50\%). The
upper scale represents the orbital phase based on the ephemeris given by
\citetext{705}. The first panel shows the light curve, rebinned at 283 sec,
from one of the three MECS aboard \B. The second panel shows the 384 sec
HPGSPC light curve, while in the third panel the 500 sec PDS data are
displayed. In both last cases binning times have been chosen as an integer
multiple of the rocking stay time.}
\label{light_curves}
\end{figure}

In Fig.~\ref{light_curves} we show the average light curve of \V\ as observed
by the three operative NFIs instruments. Data were telemetred in direct modes,
which provide complete information on time, energy and, if available, position
for each photon. The two collimated instruments (HPGSPC and PDS) were operated
in rocking mode with a 96 sec stay time for HPGSPC, and 50 sec stay time for
PDS, in order to monitor the background along the orbit --- a complete
on-source observation, apart from SAGA passages and Earth occultations, is
possible only for PDS because of its two (mechanical) collimators. The
single-collimator HPGSPC will always show stay time gaps in its data.

The first part of the observation (about 40\% of the total net time, and marked
LOW in the figure) corresponds to one of the common intensity dips showed by
\V, explained as due to passages of the neutron star through clumpy
circumstellar material \cite{776}.

\section{Spectral Analysis}

The pulse-averaged spectrum of \V\ has been fitted separately in the energy
bands of the three operative \B\ NFIs instruments, and the results are
summarized in Table~\ref{spectral_fit}. We used XSPEC V9.0 \cite{1611} as
spectral fitting program, and we refer to its User Manual\footnote{The XSPEC
User Manual is available on-line at {\sf
http://legacy.gsfc.nasa.gov/docs/xanadu/xspec/u\_manual.html}} for the details
of the functionals used for the fits. We treated separately the two intensity
states of the source. Note that the high \cdof\ for PDS spectra is mainly due
to systematic uncertainty, below 5\% in the fit residuals.

\begin{table*}
\caption[]{\B\ NFIs spectral fits to \V\ pulse-averaged spectrum. MECS data
are fit by a single power law, plus photoelectric absorption, an emission
Gaussian line and an absorption edge. The HPGSPC spectrum is fit by a power
law with exponential cutoff plus an emission Gaussian line (the inclusion
of a CRF did not improve the fit). The PDS data have been fit with a power
law modified by a high-energy cutoff of the form $\exp[(E_c - E)/E_f]$,
plus a Lorenzian CRF.}
\label{spectral_fit}
\begin{flushleft}
\begin{tabular}{l|rll|rll|rll} \hline \noalign{\smallskip}
 \multicolumn{1}{c}{}
 & \multicolumn{3}{c|}{\bf MECS} & \multicolumn{3}{|c}{\bf HPGSPC} & \multicolumn{3}{|c}{\bf PDS} \\ \multicolumn{1}{c}{}
 & \multicolumn{3}{c|}{\bf (2--10 keV)} & \multicolumn{3}{|c}{\bf (6--40 keV)} & \multicolumn{3}{|c}{\bf (20--100 keV)} \\ 
\noalign{\smallskip} \hline\hline \noalign{\smallskip} 
LOW     & N$_{\rm H}$       & $6.5\pm 0.3$           & (10$^{22}$ cm$^{-2}$) & E$_c$             & $9.6\pmt{0.4}{0.5}$    & (keV) & E$_c$         & $35 \pm 1$           & (keV) \\
        & $\alpha$          & $0.93\pmt{0.06}{0.01}$ &                       & $\alpha$          & $-0.32 \pm 0.09$       &       & $\alpha$      & $2.2\pm 0.1$         &       \\
        & $E_{\rm Fe}$      & $6.42\pm 0.03$         & (keV)                 & E$_{\rm Fe}$      & $6.44\pmt{0.05}{0.04}$ & (keV) & E$_f$         & $17\pmt{2}{8}$       & (keV) \\
        & $\sigma_{\rm Fe}$ & $0.10\pm 0.07$         & (keV)                 & $\sigma_{\rm Fe}$ & 0.10 (fixed)           & (keV) & Depth         & $0.9\pmt{0.2}{0.01}$ &       \\
        & EW                & $110\pm 25$            & (eV)                  & EW                & $375\pm 130$           & (eV)  & Width         & $10\pmt{8}{13}$      & (keV) \\
        & $E_{\rm edge}$    & $7.6\pm 0.2$           & (keV)                 &                   &                        &       & E$_{\rm cyc}$ & $56\pm 2$            & (keV) \\
        & $\tau_{\rm edge}$ & $0.15\pm 0.03$         &                       &                   &                        &       &               &                      &       \\
        & Flux$^a$          & 0.24                   &                       & Flux$^a$          & 0.34                   &       & Flux$^a$      & 0.07                 &       \\
        & Luminosity$^b$    & 0.98                   &                       & Luminosity$^b$    & 3.6                    &       & Luminosity$^b$& 1.4                  &       \\
        & \cdof\ (dof)      & 1.048 (163)            &                       & \cdof\ (dof)      & 1.167 (127)            &       & \cdof\ (dof)  & 1.501 (29)           &       \\ 
\noalign{\smallskip} \hline \noalign{\smallskip} 
HIGH    & N$_{\rm H}$       & $0.86\pmt{0.07}{0.10}$ & (10$^{22}$ cm$^{-2}$) & E$_c$             & $9.3\pmt{0.2}{0.3}$    & (keV) & E$_c$         & $39\pm 2$            & (keV) \\
        & $\alpha$          & $0.91 \pm 0.03$        &                       & $\alpha$          & $-0.40 \pm 0.05$       &       & $\alpha$      & $2.1\pmt{0.3}{0.2}$  &       \\
        & $E_{\rm Fe}$      & $6.46 \pm 0.02$        & (keV)                 & E$_{\rm Fe}$      & $6.40\pmt{0.05}{0.04}$ & (keV) & E$_f$         & $16\pm 1$            & (keV) \\
        & $\sigma_{\rm Fe}$ & 0.10 (fixed)           & (keV)                 & $\sigma_{\rm Fe}$ & 0.10 (fixed)           & (keV) & Depth         & $1.2\pm 0.2$         &       \\
        & EW                & $115\pm 10$            & (eV)                  & EW                & $380\pm 100$           & (eV)  & Width         & $12\pm 1$            & (keV) \\
        & $E_{\rm edge}$    & $7.5\pmt{0.1}{0.2}$    & (keV)                 &                   &                        &       & E$_{\rm cyc}$ & $57\pm 1$            & (keV) \\
        & $\tau_{\rm edge}$ & $0.09 \pm 0.02$        &                       &                   &                        &       &               &                      &       \\
        & Flux$^a$          & 0.62                   &                       & Flux$^a$          & 0.64                   &       & Flux$^a$      & 0.14                 &       \\
        & Luminosity$^b$    & 2.2                    &                       & Luminosity$^b$    & 6.8                    &       & Luminosity$^b$& 2.8                  &       \\
        & \cdof\ (dof)      & 1.126 (164)            &                       & \cdof\ (dof)      & 1.720 (127)            &       & \cdof\ (dof)  & 1.969 (29)           &       \\ 
\noalign{\smallskip} \hline \noalign{\smallskip}
\multicolumn{9}{l}{{\sc Note} ---  All quoted errors represent 90\% confidence level for a single parameter.} \\
\multicolumn{9}{l}{$^a$ Total flux in each instrument energy band, in units of photons cm$^{-2}$ sec$^{-1}$.} \\
\multicolumn{10}{l}{$^b$ \begin{minipage}[t]{0.95\textwidth}
Total X--ray luminosity in each instrument energy band, in units of $10^{36}$ ergs~sec$^{-1}$, assuming a distance of 1.9 kpc
\cite{329}.\end{minipage}}
\end{tabular}
\end{flushleft}
\end{table*}

From Table~\ref{spectral_fit} it is evident that any attempt of fitting the
wide-band \V\ average spectrum with a single power law, eventually modified by
energy cutoff and/or narrow-band absorption features, will be unsuccessful,
because of the completely different spectral slopes in the three NFIs energy
ranges. Indeed, the usual --- for an X--ray pulsar --- power law plus
high-energy cutoff \cite{303} gives a very poor fit to the 2--100 keV spectra,
with $\cdof\sim 3$ and 9 for the two intensity states \cite{1564}. A fit with a
different continuum, namely a power law modified by the so called Fermi--Dirac
cutoff \cite{1584}, gave an unacceptable fit, too.

Therefore we used for the continuum a spectral law that allowed the flattening
of the spectrum observed in the $\sim 10$--30 keV range. In particular we found
that two power laws modified at high energy by the {\em same\/} exponential
cutoff were able to describe the 2--100 keV spectra, expecially the LOW state.
This spectral functional has been chosen as a compromise between a reasonable
good fit and a simple law. It is also the same continuum --- the so-called NPEX
--- used by Mihara in fitting the continuum of X--ray pulsars \cite{1547}.

\begin{figure*}
%\picplace{7cm}
\vspace{7cm}
\centerline{\includegraphics{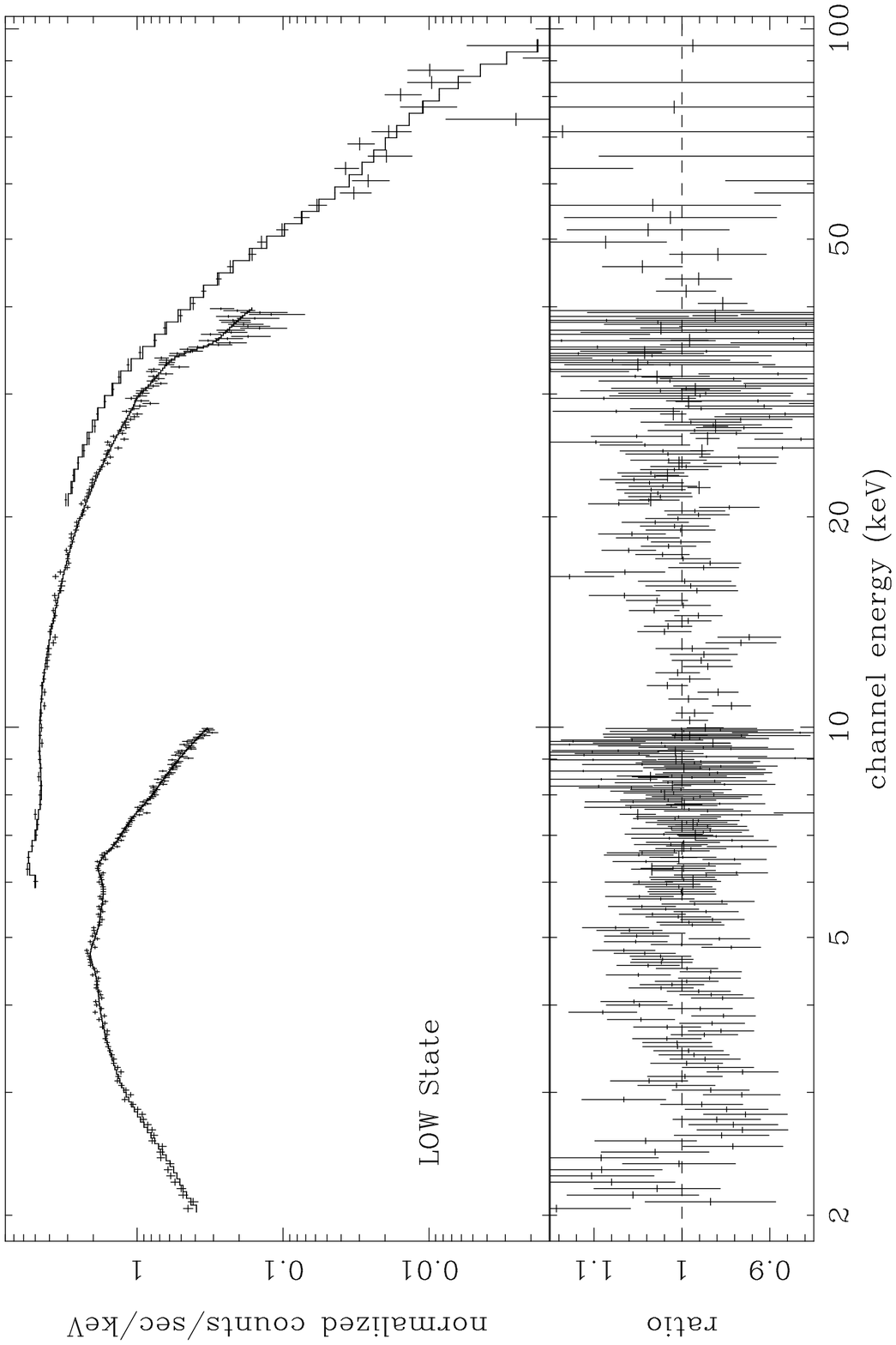}
\includegraphics{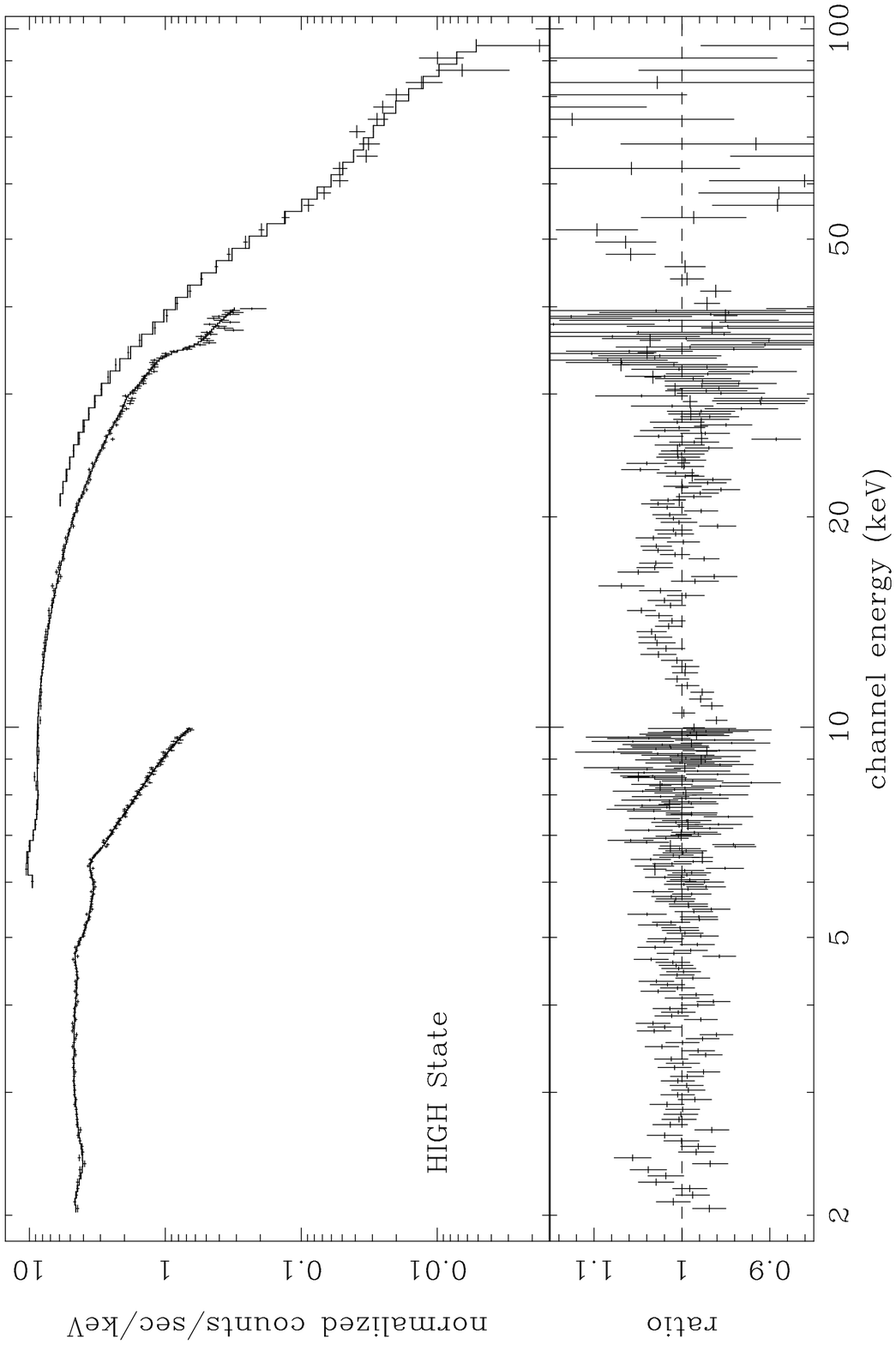}}
\caption[]{\B\ \V\ pulse-averaged spectra in the two intensity states: LOW on
the left and HIGH on the right. The continuum has been modelled by two power
laws plus the same exponential cutoff. We added an emission line, an absorption
edge and a Lorenzian cyclotron resonance at $\sim 57$ keV. Note the dramatic
change in the low energy absorption.}
\label{average}
\end{figure*}

By including a Gaussian line in emission at $\sim 6.4$ keV and an absorption
edge at $\sim 7.7$ keV we obtained a reasonable good fit, with \cdof = 1.3 (331
dof) for LOW state, and \cdof = 2.8 (332 dof) for HIGH state. From the form of
the residuals at high energy we were led to add CRFs in the spectrum
description. Other authors \cite{407,1538} have considered the fundamental
cyclotron resonance at $\sim 27$ keV, but our high energy data from both HPGSPC
and PDS do not show it. Therefore we added to our model a CRF at $\sim 57$.
Both a Lorenzian \cite{1547} or a Gaussian in absorption \cite{1172} model gave
a quite good fit, expecially for the LOW state. The fit results are summarized
in Table~\ref{total_fit} and the fits for the two intensity states are shown in
Fig.~\ref{average}. The inter-calibration between the NFIs is quite good,
expecially at the light of this early calibration status\footnote{Updated
information on the calibration status are available at {\sf
http://www.sdc.asi.it/software/}}. With MECS2 as reference, normalization is
0.98 for HPGSPC, and 0.79 for PDS.

\begin{table}
\caption[]{Fit results on the wide-band \B\ \V\ pulse-averaged spectra in the
two intensity states. Also an absorption Gaussian at $57\pm 1$ keV, with width
$22\pm 1$ keV FWHM, and Equivalent Width $30\pm 3$ keV fits well the CRF. All
quoted errors represent 90\% confidence level for a single parameter.}
\label{total_fit}
\begin{flushleft}
\begin{tabular}{lll}
\hline \noalign{\smallskip}
 & \multicolumn{1}{c}{\bf LOW State} & \multicolumn{1}{c}{\bf HIGH State} \\
\noalign{\smallskip} \hline \noalign{\smallskip} 
N$_{\rm H}$ (10$^{22}$ cm$^{-2}$) & $5.6\pmt{0.2}{0.3}$ & $0.7\pm 0.1$   \\
$\alpha_1$                        & $-2.1\pm 0.5$  & $-1.1\pm 0.2$  \\
$I_1\ ^a$                         & $(5\pm 2)\times 10^{-4}$ & $(4\pm 3)\times 10^{-3}$ \\
$\alpha_2$                        & $0.34\pm 0.11$ & $0.26\pm 0.07$ \\
$I_2\ ^a$                         & $0.13\pm 0.01$ & $0.34\pm 0.03$ \\
$E_c$ (keV)                       & $9.6\pm 1.8$   & $9.2\pm 0.8$   \\
\noalign{\smallskip} \hline
$E_{\rm Fe}$ (keV)                & $6.42\pm 0.04$ & $6.45\pm 0.02$ \\
$\sigma_{\rm Fe}$ (keV)           & 0.10 (fixed)   & 0.10 (fixed)   \\
$I_{\rm Fe}$ (Ph cm$^{-2}$ s$^{-1}$) & 
                 $(4.2\pmt{0.7}{0.9})\times 10^{-3}$ &
                 $(7.6\pmt{0.9}{0.6})\times 10^{-3}$ \\
$E_{\rm edge}$ (keV)              & $7.7\pmt{0.1}{0.2}$ & $7.4\pmt{0.1}{0.2}$ \\
\noalign{\smallskip} \hline \noalign{\smallskip}
$E_{\rm cyc}$ (keV)               & $53\pmt{1}{2}$ & $54.4\pmt{0.2}{1.5}$  \\
Depth                             & $1.5\pmt{0.6}{0.7}$ & $1.3\pmt{0.1}{0.4}$   \\
Width (keV)                       & $20\pmt{7}{4}$ & $17\pmt{2}{3}$      \\
\cdof\ (dof)                      & 1.198 (330) & 1.797 (330)    \\
\noalign{\smallskip} \hline \noalign{\smallskip}
\multicolumn{3}{l}{$^a$ Photons cm$^{-2}$ sec$^{-1}$ keV$^{-1}$ at 1 keV.}
\end{tabular}
\end{flushleft}
\end{table}

Our results on the Iron line energy at 6.4 keV are in agreement with those
found by other experiments. A more detailed discussion on the properties of
the Iron line, with emphasis on its pulse-phase variation, will be
discussed in a separate paper.

\begin{figure}
%\picplace{6cm}
\epsfxsize=0.5\textwidth   \epsffile{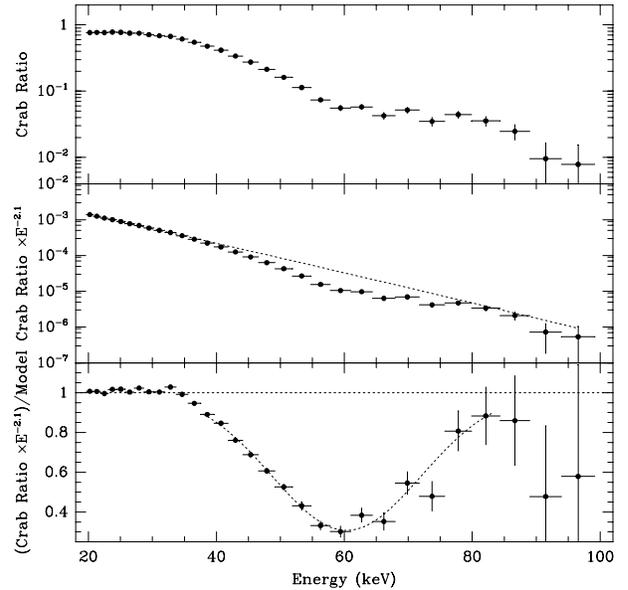}
\caption[]{{\em Upper panel\/}: Ratio between the PDS \V\ and Crab spectra.
Note the change of slope at $\sim 57$ keV. {\em Middle panel\/}: the \V/Crab
ratio multiplied by E$^{-2.1}$, the functional form of the Crab spectrum.
Deviations from the continuum (dotted line) occur for E $\ga 35$ keV. {\em
Lower panel\/}: ratio between the (\V/Crab ratio $\times$ E$^{-2.1}$) and the
functional form of the \V\ continuum as derived from the wide-band fit. A
Gaussian fit to the CRF (dotted line) is also shown.}
\label{crab-ratio}
\end{figure}

\section{Discussion}

\subsection{The X--ray continuum}

This is the first time that we have three (quite different) X--ray detectors
that {\em simultaneously\/} observed \V\ in a so wide energy range. From the
single NFIs spectra we can extrapolate that the wide-band spectrum will have a
complex form: up to $\sim 10$ keV it is very well described by a simple power
law (MECS data). Then the spectrum begins to flatten: first at $\sim 10$ keV
(cutoff in HPGSPC data) and then more pronouncedly after $\sim 35$ keV (cutoff
in PDS data). These variations are quite in agreement with the change observed
in the \V\ pulse profile, from a five-peaked to a two-peaked shape just in the
10--30 keV range (see {\em e.g.,\/} Orlandini 1993). \nocite{201}

The NPEX functional that we used in the wide-band spectral fitting is surely a
crude approximation of this behaviour, but it describes the data very well for
the LOW state, and reasonably well for the HIGH state. It is important to
stress how the choice of the functional describing the continuum has dramatic
effects on the line characteristics. Indeed, a fit to our data with a power law
modified by a high energy cutoff of the type described by \citetext{303} was
not able to well describe both the $\sim 27$ and $\sim 57$ keV CRFs
\cite{1564}.

\subsection{Cyclotron resonance features}

The \V\ pulse-averaged spectrum shows all the features present in a typical
X--ray binary pulsar spectrum, namely an Iron emission line with its absorption
edge, and signatures of cyclotron resonance. While the Iron line feature is
clearly determined in the spectrum, there are ambiguities in the position of
the fundamental CRF. Indeed, a fit to the wide-band \B\ spectrum with an NPEX
and two CRFs, with the fundamental constrained in the 10--40 keV range,
produced a cyclotron energy at $\sim 24$ keV and a \cdof\ comparable with those
shown in Table~\ref{total_fit}, although the CRF depth was extremely low. An
F--test was not successful in discriminating between the two hypothesis, with a
probability of chance improvement of 98\% for LOW State and 42\% for HIGH
State.

But there are some arguments against the low energy CRF: first, the HPGSPC
would surely be successful in its detection if it had the same characteristics
as in \citetext{1547}, while the inclusion of such feature did not improve the
fit. Second, a pulse-phase spectroscopy on the PDS data alone shows again that
a $\sim 27$ keV CRF is not detected, but only the $\sim 57$ keV CRF fits well
the data \cite{1561}. Also a ratio between PDS spectra taken during the pulse
peaks and those taken during the valleys did not show any variation at $\sim
27$ keV, but only at $\sim 57$ keV.

But the most striking result has been obtained by performing the ratio
between the PDS \V\ and Crab count rate spectra. This ratio has the
advantage of minimizing the effects due to the detector response and
uncertainties in the calibrations. As we can see from the upper panel in
Fig.~\ref{crab-ratio}, there is an evident change of slope at $\sim 57$
keV. The same ratio performed on the Hercules X--1 spectrum shows a change
of slope at $\sim 40$ keV, just its well known cyclotron energy
\cite{1583}.

To better enhance the effect due to CRFs, we multiplied the \V/Crab Ratio by
the functional form of the Crab spectrum, {\em i.e.\/} a simple power law with
index equal to 2.1. In this way we point out deviations of the \V\ spectrum
from its continuum without making any assumption on its form. The result is
shown in the second panel of Fig.~\ref{crab-ratio}, where it is evident that
deviations occur only at E $\ga 35$ keV.

Finally, in the lower panel we show the ratio between the previous function and
the \V\ continuum functional, with $\alpha_1$, $\alpha_2$, and $E_c$ taken from
Table~\ref{total_fit}. In this way we enlarge all the effects due to line
features, although we introduce a model dependence.  We can clearly see the
absorption feature at $\sim 60$ keV, that we fit between 40 and 80 keV with
both a Gaussian (shown in the figure) and a Lorenzian. The results are shown in
Table~\ref{crf_fit} and seem to indicate a preference for the Gaussian shape.

\begin{table}
\caption[]{Fit results on the \V\ CRF. Errors represent 90\% confidence level.}
\label{crf_fit}
\begin{flushleft}
\begin{tabular}{lll}
\hline \noalign{\smallskip}
 & \multicolumn{1}{c}{\bf Gaussian} & \multicolumn{1}{c}{\bf Lorenzian} \\
\noalign{\smallskip} \hline \noalign{\smallskip} 
$E_{rm cyc}$ (keV)       & $60.4\pm 0.7$ & $59.2\pm 0.6$ \\
$\sigma_{\rm cyc}$ (keV) & $11.6\pm 0.5$ & $22\pm 1$     \\
\cdof\ (dof)             & 1.275 (12)    & 6.662 (12)    \\
\noalign{\smallskip} \hline
\end{tabular}
\end{flushleft}
\end{table}

Because of these considerations we will interpret the $\sim 57$ keV CRF as the
fundamental cyclotron resonance. This corresponds to a surface magnetic field
strength of $4.9\times 10^{12}$ Gauss, in agreement with theoretical
considerations about a high magnetic field in \V\ \cite{1099}. 

At the cyclotron resonance frequency $\omega_c$, electrons at rest absorb
photons of energy $\hbar\omega_c$. For moving, thermal electrons the Doppler
broadening $\Delta\omega_D$ is predicted to be \cite{1164}

\begin{equation}
\Delta\omega_D = \omega_c \left( \frac{2kT_e}{m_ec^2} \right)^{1/2}
  |\cos\theta| 
\label{eq:deltaD}
\end{equation}
where $kT_e$ is the electron temperature, and $m_ec^2$ is the electron rest
mass. The angle $\theta$ measures the direction of the magnetic field with
respect to the line of sight. Outside the range $\omega_c\pm\Delta\omega_D$ the
cyclotron absorption coefficient decays exponentially, and other radiative
processes become important. From Eq.~\ref{eq:deltaD} and the CRF parameters
given in Table~\ref{total_fit} we obtain a lower limit to the electron
temperature of $\sim 8$ keV, in fair agreement with the calculations of
self-emitting atmospheres of \citetext{304}. A $\sim 27$ keV CRF would
corresponds to an electron temperature, assuming the same FWHM, of $\sim 34$
keV: a bit too high. On the other hand, by imposing $kT_e\sim 8$ keV would
correspond to a CRF FWHM of $\sim 10$ keV that should be clearly visible in our
spectra.

Finally, the $\sim 57$ keV CRF agrees with the empirical positive correlation
found between the magnetic field strength and the spectral hardness, whether
defined in terms of hardness ratio \cite{1286} or high energy cutoff \cite{61}.

\section{Conclusions}

The wide-band average spectrum of \V\ as observed by \B\ appears to be quite
complex, showing a progressive change of slope: first at $\sim 10$ keV and then
at $\sim 35$ keV, just in the same interval where the \V\ pulse profile changes
from a five-peak to a two-peak shape. This is clearly visible in the single
NFIs spectra, described by power laws with significantly different indices. We
approximate this behaviour with a double power law modified by an exponential
cutoff (the so called NPEX model). From the form of the residuals we added a
cyclotron resonance in the fitting functional: the \B\ data do not require a
CRF at $\sim 27$, as found in earlier works, but only at $\sim 57$ keV. A
possible explanation could be an intrinsic variability in the cyclotron
resonance, or an effect due to the change of slope of the continuum, not
previously taken into account. A better description of the continuum, beyond
the scope of this paper is surely needed.

\begin{acknowledgements}
The authors wish to thank the \B\ Scientific Data Center staff for their
support during the observation and data analysis. This research has been funded
in part by the Italian Space Agency.
\end{acknowledgements}

\end{document}